\documentclass{revtex4}

\usepackage{graphicx}
\preprint{P3-19}
\preprint{CERN-TH/2001-255}

\begin{document}
\bibliographystyle{apsrev}
\hfill\parbox{8cm}{\raggedleft CERN-TH/2001-255 \\ hep-ph/0110227 \\
Snowmass-2001/P3-19 \\
}

\title{Theoretical Uncertainties in Sparticle Mass Predictions}
\author{B.C. Allanach}
\email[]{benjamin.allanach@cern.ch}
\homepage[]{http://allanach.home.cern.ch/allanach/}
\affiliation{TH Division, CERN, Geneva 23, CH 1211 Switzerland}
\date{\today}

\begin{abstract}
We contrast the sparticle spectra obtained from 
three modern publicly available codes
along model lines in minimal supersymmetric standard model (MSSM) parameter
space. 
From this we gain an idea of the uncertainties involved with sparticle spectra
calculations.
The differences in predicted sparticle masses are typically at the several
percent-level. In the focus-point scenario, there are differences of $30\%$ in
the weak gaugino masses. These uncertainties need to be reduced in 
order to obtain accurate information about fundamental models of supersymmetry
breaking. 
\end{abstract}

\maketitle

%Supersymmetric phenomenology is notoriously complicated. 
%Even if one assumes the particle spectrum of the minimal supersymmetric
%standard model (MSSM),
%fundamental patterns of supersymmetry (SUSY) breaking are numerous.
%It seems that there is currently nothing to strongly favor one particular scenario
%above all others.
%Collider signatures typically rely upon identifying decay products
%of produced sparticles through cascade decay chains.
%The resulting signatures of different scenarios of SUSY breaking are not only
%highly dependent upon the scenario that is assumed, but also upon any model
%parameters~\cite{atlastdr}. 
%As a supersymmetry breaking parameter is changed, the ordering of
%sparticle masses can change, switching various sparticle decay branches on
%and off. 

MSSM phenomenology is so complicated
that  
studies of the ability of future colliders to search for and measure
supersymmetric 
parameters has focused on isolated `bench-mark' model
points~\cite{Allanach:2000kt,Battaglia:2001zp,atlastdr}. This approach,
while 
being a start, is not ideal because one is not sure how many of the
features used in the analyses will apply to other points of parameter space.
In an attempt to cover more of the available parameter space, the 
{\em Direct Investigations of SUSY Subgroup} of {\em SNOWMASS 2001} has
proposed eight bench-mark model lines for study.

The lines have been defined to have the spectrum output from the {\tt ISASUGRA}~
program (part of the {\tt ISAJET7.51} package~\cite{Baer:1999sp}) for $m_t=175$ GeV.
{\tt ISASUGRA}~ solves the MSSM normalization group equations subject to boundary
conditions at low energy (measured Standard Model couplings and constraints
from consistent electroweak symmetry breaking) and 
a higher energy scale (the theoretical boundary condition on the soft mass
parameters).
Knowledge of the uncertainties will be important when data
is confronted with theory, i.e. when information upon a high-energy SUSY
breaking sector is sought from low-energy data.
Here, we intend to investigate the theoretical uncertainties in sparticle
mass determination. To this end, we contrast the sparticle masses predicted by
three modern publicly available supported codes: {\tt ISASUGRA}, {\tt
SOFTSUSY}~\cite{Allanach:2001kg} and {\tt SUSPECT}~\cite{Djouadi:1998di}. 

Each of the three packages calculates sparticle masses in a similar way, but
with different approximations~\cite{stefano}. In each of the model line
scenarios, we calculate the fractional difference for some sparticle $s$
\begin{equation}
f_s^{\mbox{\tiny CODE}} = \frac{m_s^{\mbox{\tiny ISASUGRA}} - m_s^{\mbox{\tiny
CODE}}}{m_s^{\mbox{\tiny ISASUGRA}}},
\end{equation}
where {\small CODE}~refers to {\tt SOFTSUSY1.2}, or {\tt SUSPECT2.0}. $f_s^{\mbox{\tiny CODE}}$ 
then gives the 
fractional difference of the mass of sparticle $s$ between the predictions of
{\small CODE} and {\tt ISASUGRA}. A positive value of $f_s^{\mbox{\tiny CODE}}$ then
implies that $s$ is heavier in {\tt ISASUGRA}~ than in {\small CODE}. 

We focus upon model lines in scenarios which are currently supported by all
three packages, i.e. supergravity mediated supersymmetry breaking (mSUGRA).
At a high unification scale $M_{GUT}\equiv 1.9 \times 10^{16}$, 
the soft-breaking scalar masses are set to be all equal to $m_0$, the
universal scalar trilinear coupling to $A_0$ and each gaugino mass $M_{1,2,3}$ 
is set. $\tan \beta$ is set at $M_Z$. 
The three choices of model lines are displayed in Table~\ref{P3-19:tabmodels}.
 \begin{table}
 \caption{Model lines in mSUGRA investigated here. $m_t=175$ GeV, $M_{GUT}=1.9 \times 10^{16}$ GeV and
$\alpha_s(M_Z)^{\overline{MS}} = 0.119$ are used.}
 \label{P3-19:tabmodels}
 \begin{tabular}{|c|ccccccc|} \hline
Model line & $\tan \beta$ & $A_0$ & $M_1$ & $M_2$ & $M_3$ & $m_0$ & sgn$\mu$
\\ \hline
A          & 10 & -0.4$M_{1/2}$ & $M_{1/2}$ & $M_{1/2}$ & $M_{1/2}$ &
0.4$M_{1/2}$ & + \\
B & 10 & 0 & 1.6$M_2$ & $M_2$ & $M_2$ & $M_2/2$ & + \\
F & 10 & 0 &  $M_{1/2}$ & $M_{1/2}$ & $M_{1/2}$ & $2M_{1/2}+800$ GeV & + \\ \hline
 \end{tabular}
 \end{table}
Model line A displays gaugino mass dominance, ameliorating the SUSY flavor
problem. Model line B has non-universal gaugino masses and 
model line F corresponds to focus-point supersymmetry~\cite{Feng:1999mn},
close to the electroweak symmetry breaking boundary.

\begin{figure}
\unitlength=1in
\begin{picture}(6,2.5)
\put(0,0){\includegraphics[scale=0.75]{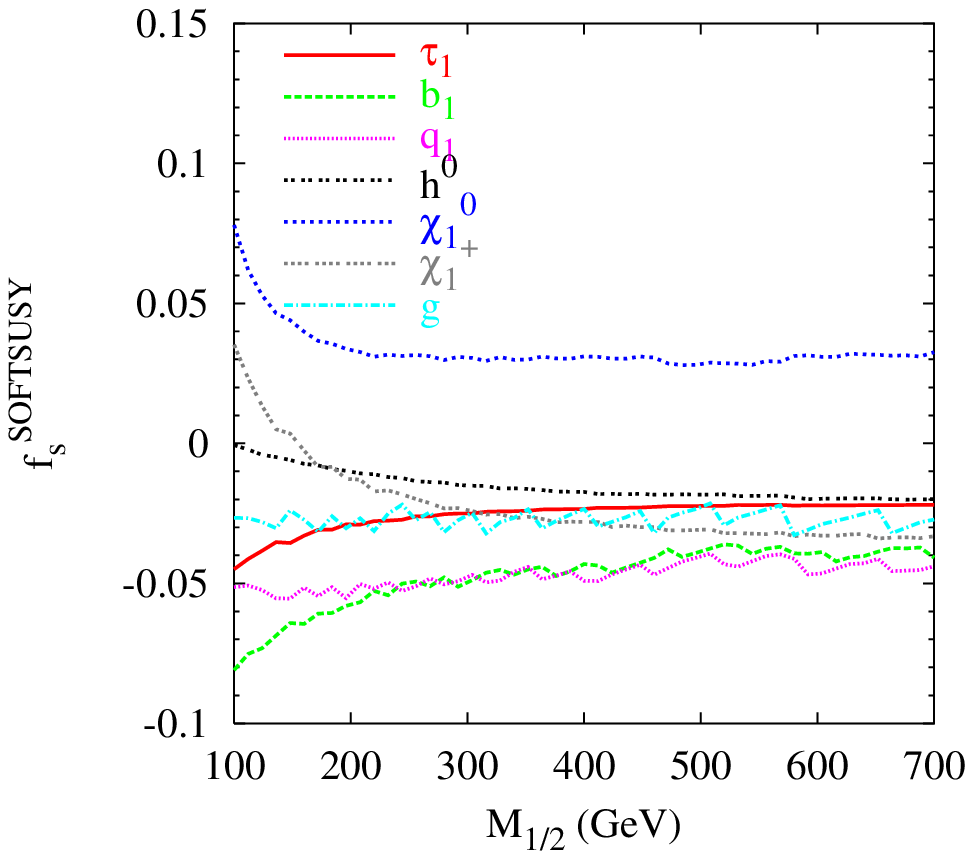}}
\put(3,0){\includegraphics[scale=0.75]{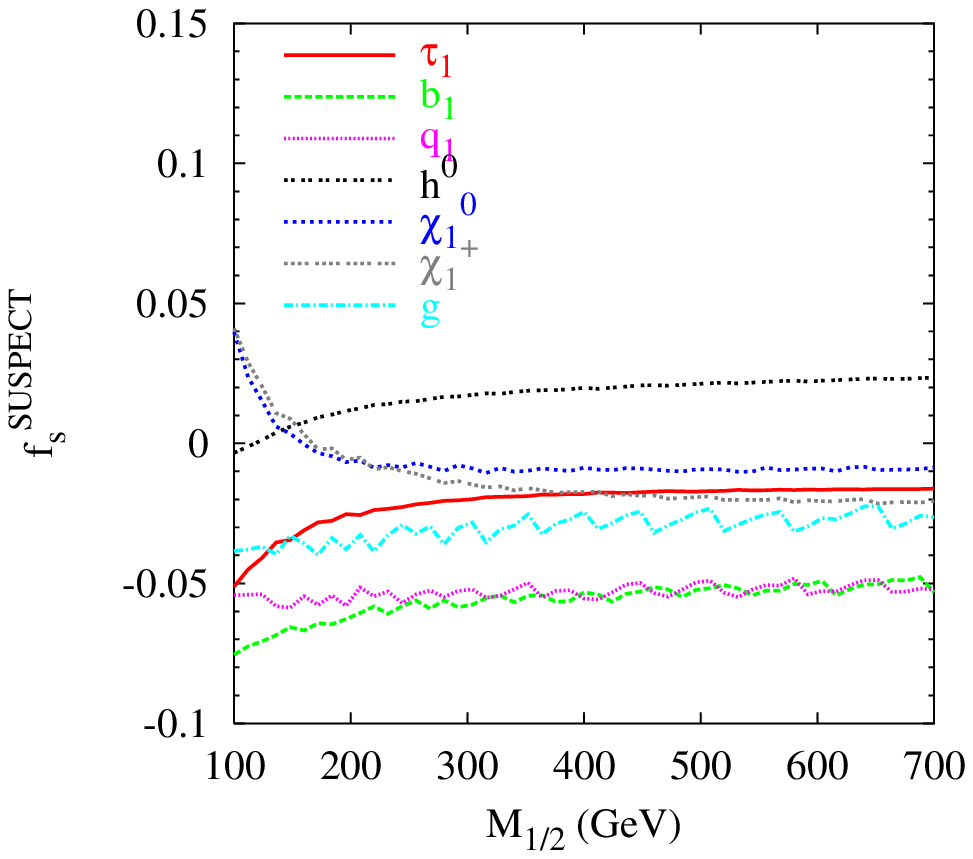}}
\put(0.,2.4){(a)}
\put(3.0,2.4){(b)}
\end{picture}
 \caption{Fractional differences between the spectra predicted for mSUGRA
model line A}
 \label{P3-19:fig:modelA}
 \end{figure}
\begin{figure}
\unitlength=1in
\begin{picture}(6,2.5)
\put(0,0){\includegraphics[scale=0.75]{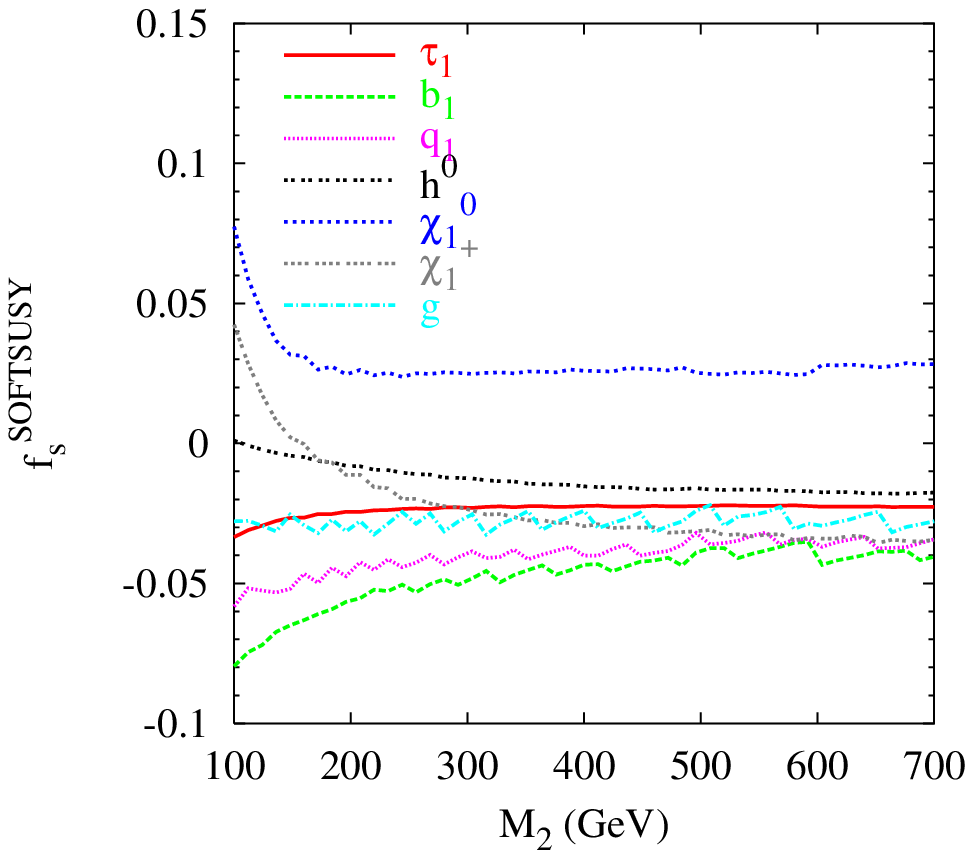}}
\put(3,0){\includegraphics[scale=0.75]{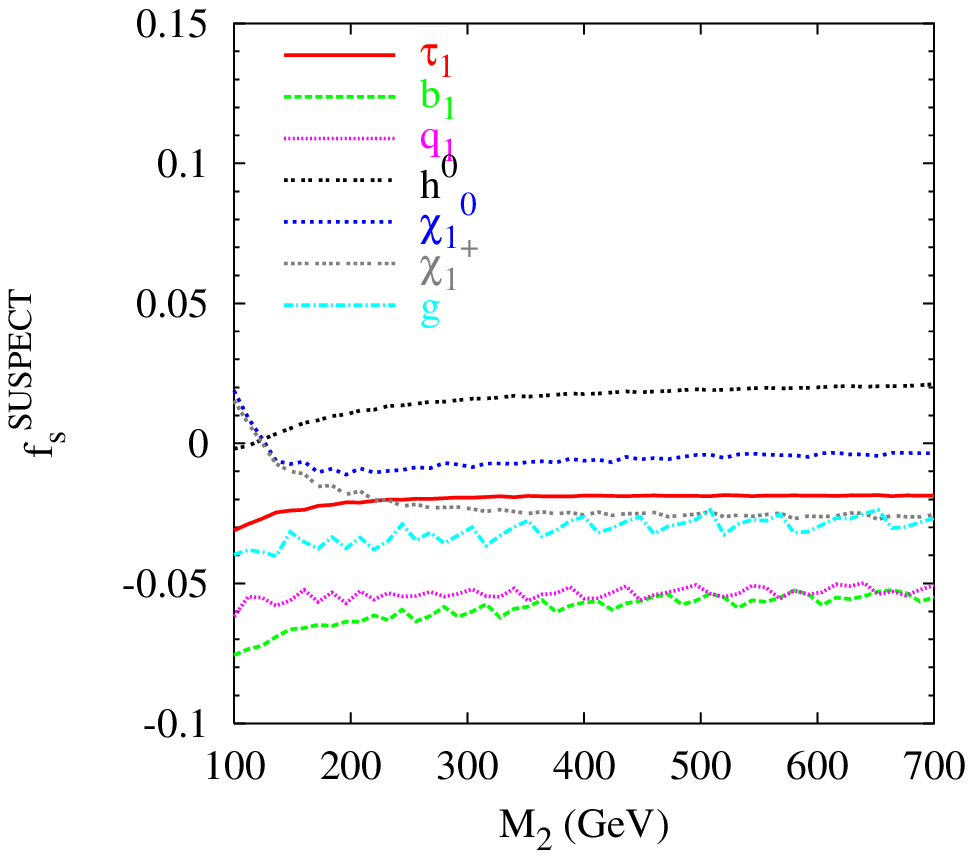}}
\put(0.,2.4){(a)}
\put(3.0,2.4){(b)}
\end{picture}
 \caption{Fractional differences between the spectra predicted for
model line B}
 \label{P3-19:fig:modelB}
 \end{figure}

We pick various sparticle masses that show a large difference in their
prediction between the
three calculations. For model line A,
Fig.~\ref{P3-19:fig:modelA}a shows
$f_{\tau_1,b_1,q_1,h^0,\chi_1^0,\chi_1^+,g}^{\mbox{\tiny SOFTSUSY}}$
(the lightest stau, sbottom, squark, neutral Higgs, neutralino, chargino and
gluino mass difference fractions respectively). Fig.~\ref{P3-19:fig:modelA}b
shows 
the equivalent results for the output of {\tt SUSPECT}. 
Model line B differences are shown in Fig.~\ref{P3-19:fig:modelB}.
Jagged curves in the figures are a result of numerical error in the {\tt
ISASUGRA} calulcation, and are at the
1$\%$ level for squarks, gluinos and the lightest neutralino. 
The stau,
lightest Higgs and lightest chargino do not display any appreciable numerical
error. 
We have checked that
{\tt SOFTSUSY} and {\tt SUSPECT} do not have any numerical error that is
detectable by eye for any of the sparticles.

Figs.~\ref{P3-19:fig:modelA},\ref{P3-19:fig:modelB} share some common
features. 
In general, the largest discrepancies occur for low $M_{1/2}$, where the
super-particle spectrum is lightest.
The gluino and squark masses are consistently around 5$\%$ lower in
{\tt ISASUGRA}~ than the other two codes, which agree with each other to better than 
$1\%$. 
We note here that this uncertainty is not small,
a $5\%$ error on the gluino mass at $M_{1/2}=700$ GeV in model A corresponds
to an error on the predicted gluino mass of 75 GeV, for example.
The lightest neutralino mass predicted by {\tt SUSPECT} is in agreement with
{\tt ISASUGRA}~ to $2\%$, whereas {\tt SOFTSUSY} predicts this neutralino to be lighter by
another $2\%$ or so. The Higgs masses also agree to above 2$\%$ between each
code and {\tt ISASUGRA}, but {\tt SOFTSUSY} and {\tt SUSPECT} have opposite
sign 
differences. This is probably due to the fact that {\tt SOFTSUSY} uses a
FeynhiggsFast calculation of the neutral Higgs
masses with important two-loop effects added~\cite{Heinemeyer:1999be}, which
predicts masses that tend to be higher 
than the one-loop calculation (as used in {\tt ISASUGRA} or {\tt SUSPECT}).
The stau mass also looks fairly consistent between {\tt SUSPECT} and {\tt SOFTSUSY} (2\%
lighter than {\tt ISASUGRA}), but this could conceivably be due to their one-loop
running of scalar masses. The lightest neutralino mass shows large deviations
between all codes up 
to 5$\%$, and this is not currently understood. 

\begin{figure}
\unitlength=1in
\begin{picture}(6,2.5)
\put(0,0){\includegraphics[scale=0.75]{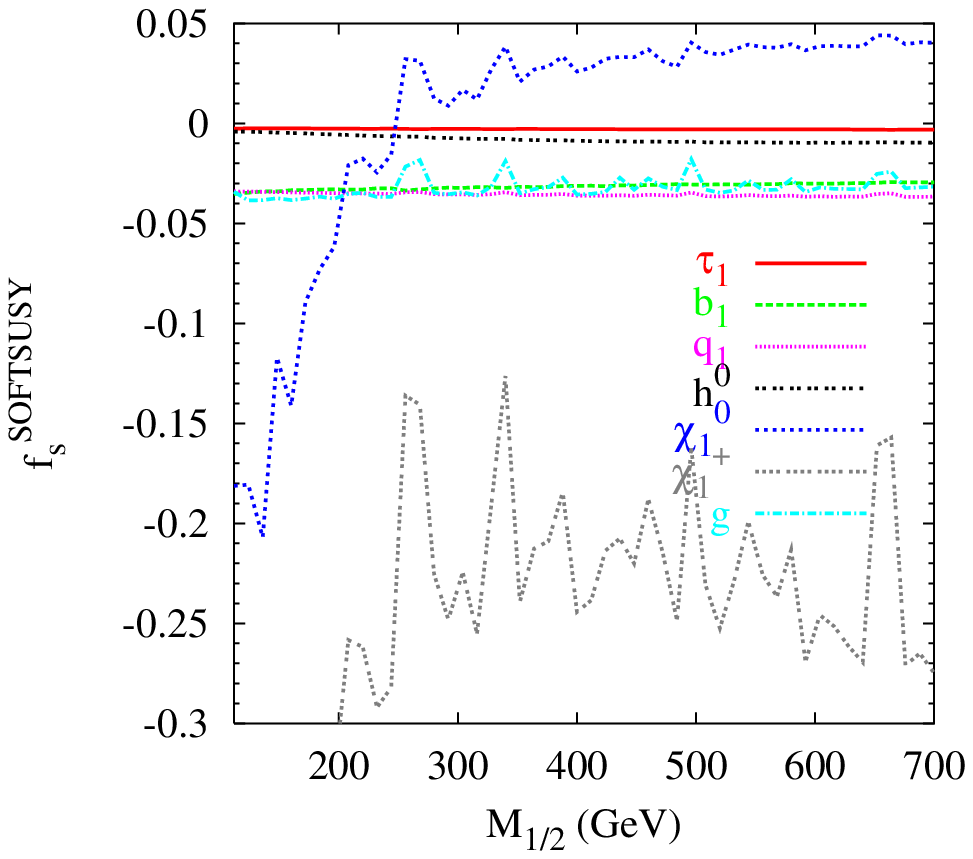}}
\put(3,0){\includegraphics[scale=0.75]{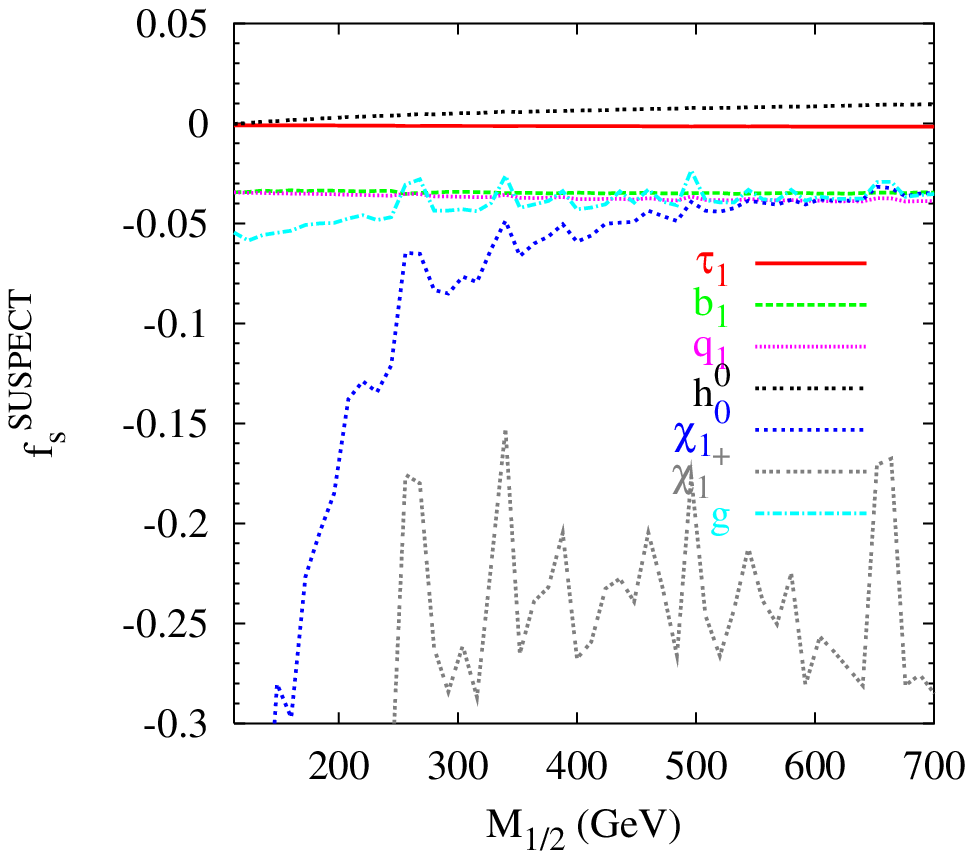}}
\put(0.,2.4){(a)}
\put(3.0,2.4){(b)}
\end{picture}
 \caption{Fractional differences between the spectra predicted for 
model line F}
 \label{P3-19:fig:modelF}
 \end{figure}

The focus-point scenario (model line F) is displayed in
Fig.~\ref{P3-19:fig:modelF}.
The overall view of spectral differences is similar to that in model lines A
and B except for the masses of the lightest chargino and neutralino.
They display large 10-30$\%$ differences. The large spikes come from numerical
errors in the {\tt ISASUGRA} calculation. 
In focus point supersymmetry, the bilinear Higgs mass parameter $\mu$ is close
to zero and is very sensitive to threshold corrections to
$m_t$~\cite{Allanach:2000ii}. For small $\mu < M_Z$, the lightest chargino and
neutralino masses become sensitive to its value. The predicted
value of $\mu(M_Z)$ differs by 50$\%$-100$\%$ between all of the three code's
output.
Only a few of the threshold corrections to $m_t$ are included 
in the {\tt ISASUGRA}~ calculation, whereas {\tt SOFTSUSY}, for example,
includes all one-loop corrections with sparticles in the loop. {\tt SUSPECT}
also adds many of 
the sparticle loop corrections to $m_t$.

In order to discriminate high energy models of supersymmetry
breaking, it will be necessary to have better than 1$\%$ accuracy in
both the experimental {\em and} theoretical determination of superparticle
masses~\cite{agq}. 
An alternative bottom-up approach~\cite{Blair:2000gy} is to evolve soft
supersymmetry breaking
parameters from the weak scale to a high scale once they are `measured'. 
The parameters of the high-scale theory are then inferred, and theoretical
errors involved in the calculation will need to be minimised. 
We note that possible future linear colliders could determine some
sparticle masses at the per-mille level~\cite{Aguilar-Saavedra:2001rg}.
An increase in accuracy of the theoretical predictions of sparticle masses by
about a factor 10 will be necessary. 

\begin{acknowledgments}
I would like to warmly thank Werner Porod for his comments on the work.
\end{acknowledgments}

\end{document}